\documentclass[3p,twocolumn,a4paper,authoryear]{elsarticle}
\usepackage{graphicx}
\usepackage{color}
\usepackage[colorlinks=true]{hyperref}
\usepackage{amsmath}
\usepackage{amssymb}
\usepackage{color}
\usepackage{natbib}

\journal{Journal of Theoretical Biology (accepted)}

\begin{document}
\begin{frontmatter}
\title{Waves of seed propagation induced by delayed animal dispersion}

%\title{Effect of a delay on the velocity of vegetation propagation mediated by animals}

%Effect of delayed dispersion on the velocity of seed propagation mediated by animals \\
%Effect of delayed dispersion on the velocity of seed propagation \\
%Effect of delays on the velocity of seeds dispersed by animals \\
%Effect of delayed dispersion on the velocity of animal mediated seed dispersion

\author[bariloche]{Laila D. Kazimierski}
\ead{laila.kazimierski@cab.cnea.gov.ar}

\author[bariloche]{Marcelo N. Kuperman}
\ead{kuperman@cab.cnea.gov.ar} 

\author[ifca,ifisc]{Horacio S. Wio}
\ead{wio@ifca.unican.es}

\author[bariloche]{Guillermo Abramson}
\ead{abramson@cab.cnea.gov.ar}

\address[bariloche]{Centro At\'omico Bariloche, CONICET and Instituto Balseiro, R8402AGP San Carlos de Bariloche, Argentina}

\address[ifca]{Instituto de F\'{\i}sica de Cantabria (UC\&CSIC), Avda.\ de los Castros, s/n, E-39005 Santander, Spain}

\address[ifisc]{Instituto de F\'{\i}sica Interdisciplinar y Sistemas Complejos (UIB\&CSIC), Campus Universitat de les Illes Balears, E-07122 Palma de Mallorca, Spain}

\date{Received: date / Revised version: date}
% The correct dates will be entered by Springer

\begin{abstract}
We study a model of seed dispersal that considers the inclusion of an animal disperser moving 
diffusively, feeding on fruits and transporting the seeds, which are later deposited and capable of 
germination. The dynamics depends on several population parameters of growth, decay, harvesting, 
transport, digestion and germination. In particular, the deposition of transported seeds at places 
away from their collection sites produces a delay in the dynamics, whose effects are the focus of 
this work. Analytical and numerical solutions of different simplified scenarios show the existence 
of travelling waves. The effect of zoochory  is apparent in the increase of the velocity 
of these waves. The results support the hypothesis of the relevance of animal mediated seed 
dispersion when trying to understand the origin of the high rates of vegetable invasion observed 
in real systems.
\end{abstract}

%\maketitle
\end{frontmatter}

\section{Introduction}
%Spatially distributed processes have been recognized as essential mechanisms in defining 
%the structure and dynamics of populations. 
One of the most relevant processes governing the 
dynamics of spatial patterns in plant populations is seed dispersal. There are numerous examples showing that the geographical advance of 
vegetation is much faster that what can be predicted from the short seed 
dispersion distance provided only by physical means, without intervening animal agents. 
Indeed, observed and recorded rates of invasion (and velocity of migration) of plants are sometimes more 
than one order of magnitude greater than expected. The origin of this discrepancy is rooted in 
a combination of multiple effects among which we can mention the disperser agents and  
the seed morphology. The disperser that acts as vector for seed dissemination can be 
abiotic or biotic and their relative importance is still a matter of study. In some cases, where the action of small animals fails to provide an explanation, there are structured seeds that profit from  efficient wind dispersal \citep{bullock17, tamme14}. For example, a 
thorough analysis of seed dispersal in the tropical forest shows that mean dispersal distances due 
to wind intervention are comparable with those where mammals or birds are involved \citep{muller08}. 
Nevertheless, it is important to seek a partial answer of the dilemma by trying to 
characterize the long distance dispersal attributed to zoochory, since it presents unique 
aspects affecting its dynamics.

Unveiling the mechanisms of this fast propagation is not only interesting to understand 
historical processes. It is also a matter of high relevance as the ability of plants to 
propagate fast and invade larger areas is crucial for its survival within present scenarios of 
changing environment due to climate, fragmentation or invasion by competitors or predators \citep{pitelka1997,cain2000,bullock2002}.
A particular example of this phenomenon is the fast rate of post-glacial migration. 
At the beginning of the Holocene, mainly due to a change in climate conditions, there was 
a rapid shift in global vegetation \citep{reid1899,skellam1951}, which is
responsible for the  current distribution of many herbaceous plants \citep{cain1998}. 
Extrapolated migration rates during the Holocene indicate that they are not compatible with the 
measured dispersal distances. This discrepancy has been called Reid's paradox by \citet{clark1998} (after \citet{reid1899}).
Despite many years of research on seed dispersal \citep{ridley1930,murray1986,bullock2002}, 
there are still gaps in our knowledge of how seeds travel long distances.

In many temperate and tropical ecosystems, the majority of seeds dispersers of 
woody plants are frugivorous animals \citep{herrera2002}. 
For example, large grazing mammals have long been recognized as potentially important seed 
dispersers \citep{ridley1930,dore1942,welch1985,malo2000,heinken2002}  and, eventually, responsible 
for the high dispersal rates \citep{vellend03}. For these plants, seed dispersion is a 
function of animal movement and gut passage times of seeds \citep{murray1988,schupp1993}. For this 
reason, one expects that the dispersal rate and the spatial pattern of plant distribution feeds 
back into the characteristics of seed dispersal via its effects on animal movements.

Then, it is not surprising 
the continuing effort devoted to obtain  more accurate and thorough 
models unveiling the interplay of animal movement and seed dispersal. Still, there are not many mathematical models of dispersion that emphasize the enhancement of dispersal rate due to animal agents.
\citet{pakeman2001} is an example, proposing a model that 
analyses how plant migration rates vary with herbivore home range, gut survival and probability of 
consumption. His results show that the role of  herbivores with long displacements and large home 
ranges  is essential to explain high rates of dispersal in the palaeorecord. The hypothesis that 
large herbivores are the  main responsible for the dispersal of seeds can explain the observed 
advance of  woodland herbs. In the same spirit, \citet{neubert2000} shows that when the dispersal occurs through  both
long and short distance mechanisms, it is the long-distance component the one that determines 
the invasion speed. It is also known that spatial patterns can arise as the result of trophic interactions and dispersal, and a 
number of scientists have investigated this question using continuous-time growth models with 
simple (Fickian) diffusion \citep{neubert1995}. Currently these phenomena also capture the interest 
of physicists and mathematicians, who seek to provide a theoretical framework for them.

Seed dispersal has also a major influence on plant fitness 
because it determines the locations in which subsequent seedlings live or die 
\citep{howe82, schupp1993, schupp2010, wenny2001}. As such, it determines not only the ecological 
dynamics, but also plant evolution and the rates of gene flow. 
Moreover, the relationship between plants and their seed dispersers is generally of a mutualistic nature, since both 
derive some benefit from their participation: food reward exchanged 
for the service of seed transport. In general the pattern of seed dispersal and activities of their dispersers are closely related \citep{neupane2015,wenny2001}, and in many cases it is possible 
to trace a co-evolutionary natural history of both. Indeed, the biological system that 
has inspired us in this work is the mutualistic relation between the marsupial 
\textit{Dromiciops gliroides} and the parasitic mistletoe \textit{Tristerix corymbosus}, a keystone species of the Patagonian temperate forest 
\citep{amico2000,garcia2009,morales2012}. \emph{D. gliroides} is its major seed disperser, so the 
arrangement 
of future generations of plants depends on the places that are visited by the animals. 
These, in turn, are the fructifying plants that provide the animal one of their main 
resources. 

In this context, delay models of seed dispersal have been 
studied extensively \citep{hadeler2007}. For example, \citet{morita1984} and  
\citet{oliveira1994} performed thorough studies of periodic solutions of diffusion equations with 
delay, while \citet{faria1999} and \citet{freitas1997} investigate bifurcations in such problems. In this work we focus on the effects induced by the 
characteristic delay between consumption and deposition of seeds on the velocity of vegetation 
dispersal. Our model involves several aspects of the cycle of seed dispersal, in which an animal eats fruit, moves over space following certain rules, and  after 
some time deposits the seeds at a different place, where a new plant will eventually grow. We are interested in the 
description of the spatio-temporal characteristics of such a dynamics.
We approach the problem as a 
reaction-diffusion system, in which consumption of seeds and their delayed deposition after being 
transported by animals is responsible for the dispersion. We show, specifically, how the delayed deposition provided by the animals enhance the velocity of propagation of a front of vegetation.

%----------------------------------------------------------------------------------
\section{Model definition and dynamics}

For populations with overlapping generations, population size can usually be regarded as a 
continuous function of time, and an adequate mathematical tool is a set of differential equations 
\citep{murray1989}. Let us consider a three-components model: seeds in plants ($f$), being dispersed 
by animals ($u$), and deposited in an appropriate substrate ($s$). Seeds in populations $f$ and $s$ are 
immobile, while those in $u$ are carried by their transporters. 

\begin{figure}[t]
\centering
\includegraphics[width=\columnwidth, clip=true]{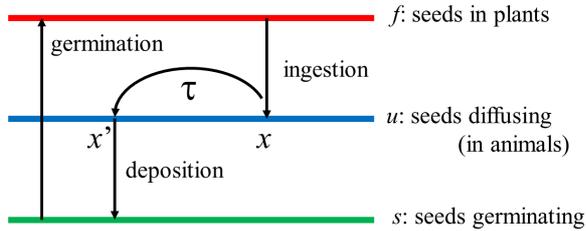}
\caption{Schematic representation of the delayed dynamics of dispersion. The lines 
represent three seed populations extended in space: $f$ are immobile seeds in fructifying plants; 
$u$ are seeds being dispersed diffusively; $s$ are seeds deposited in the substrate after a delay 
$\tau$, which eventually produce new fructifying plants.}
\label{fig:seeds}
\end{figure}

Several biological processes are mediated by interactions between these 
populations, as schematized in Fig.~\ref{fig:seeds}: 
seeds in plants are ingested and become dispersive; these are eventually deposited or defecated and 
become immobile again. Finally, with a probability of germination, the seeds grow into plants and 
start producing new seeds at the level $f$. A reasonable one-dimensional mathematical description 
of 
such a system is the following:
\begin{align}
&\frac{\partial f(x,t)}{\partial t} = F(f(x,t),s(x,t)) - I(f(x,t),u(x,t)), \label{ecf}\\
&\frac{\partial u(x,t)}{\partial t} =  I(f(x,t),u(x,t)) + D \nabla^2 u(x,t) \notag\\
    &~- \alpha\! \int^{+\infty}_{-\infty}\! 
G(x,t|x',t\!-\!\tau)\,I(f(x',t\!-\!\tau),u(x',t\!-\!\tau))\,dx',\label{ecu} 
\\
&\frac{\partial s(x,t)}{\partial t} = -g\,s(x,t) \notag \\
	&~+\alpha\! \int^{+\infty}_{-\infty}\!  
G(x,t|x',t\!-\!\tau)\,I(f(x',t\!-\!\tau),u(x',t\!-\!\tau))\,dx', 
\label{ecs}
\end{align}
that can also be formulated in higher dimensions without difficulty.

Each one of the terms in these equations represents some of the mechanisms that play a role in the 
population dynamics. $F(f,s)$ is a growth (or ripening) function of the fruits. The second term in 
Eq.~(\ref{ecf}) represents the consumption of fruit, which depends on the presence of animals 
through $u$. The same ingestion term $I(f,u)$ acts as a source of the population of moving seeds 
$u(x,t)$ in Eq.~(\ref{ecu}). 
Also, in this equation, we use a standard diffusive transport mechanism for these mobile seeds, 
with 
a coefficient $D$ and the corresponding (Gaussian) diffusion kernel that propagates from point 
$(x',t-\tau)$ to $(x,t)$ (indicated by the vertical dash in $G$): 
\begin{align}
&G(x,t|x',t-\tau) \equiv \notag\\
&G(x,x',\tau) = \exp\left(-\frac{(x-x')^2}{4D\tau}\right)(4\pi D\tau)^{-1/2}, \label{difusion}
\end{align}
whose role in the dispersion we describe below. 

The most involved term of the dynamics is the one representing the loss of mobile seeds as the 
animals deposit them. If seeds are deposited after a time $\tau$ (e.g. after the 
transit through the digestive tract of the animals), then we can propose the non-local term that 
appears in third place 
in Eq.~(\ref{ecu}): seeds are consumed at $x'$ at a time $t-\tau$, and are subsequently transported 
by the dispersion kernel $G$ up to $x$ where they are deposited at time $t$. A rate $\alpha$ takes 
into account that the process may be imperfect, with some seeds being lost and not appropriately 
transferred into the germinating population $s(x,t)$. Finally---Eq.~(\ref{ecs})---seeds deposited 
in the immobile substrate $s(x,t)$ can germinate at rate $g$ and eventually  contribute to the fruit population.

Observe that, in the absence of coupling, Eq.~(\ref{ecu}) is a reaction-diffusion equation akin to 
Fisher's equation~\citep{murray1989}. That is, with appropriate initial and boundary conditions the 
population of dispersing seeds should display a travelling wave shape, with a well defined velocity 
given by the diffusion coefficient and the linear growth rate of $I(u)$. We want to study the 
existence of similar waves in the complete coupled system and, eventually, propagating waves of 
$f(x,t)$, that is, of the fructifying plants. In order to proceed with the analysis, let us 
consider 
specific forms of the different functions involved: 
\begin{align}
F(f,s) &= r_f s(x,t)\left(k_f-f(x,t)\right), \label{efe}\\
I(f,u) &= r_u u(x,t)\left(k_u-u(x,t)\right)\frac{f(x,t)}{b+f(x,t)}, \label{ingestion}
\end{align}
where $r_i$ are growth rates and $k_i$ are carrying capacities.
Observe that Eq.~(\ref{efe}) provides a net reproduction of $f$ which is proportional to their 
source, $s$. Equation~(\ref{ingestion}), in turn, is a product of a logistic growth in $u$ 
(corresponding to the animals that disperse the seeds) and a function of $f$ that saturates as 
$f\to\infty$, indicating satiation (characterized by an additional parameter, $b$). 

The analysis of these differential equations require approximate analytical approaches as well as 
numerical solutions, which we discuss below.

%----------------------------------------------------------------------------------
\subsection{Asymptotic analysis}
Let us first consider a simplified scenario, in which the dynamics of the immobile population of 
plants $f$ occurs more slowly than that of dispersing seeds $u$. This hypothesis (to be relaxed later, showing qualitatively similar results) allows us to consider $f$ as 
a parameter, so that the ingestion term $I$ is just a function of $u(x,t)$. Let us look for 
travelling wave solutions in the usual way. Consider the change of variables to a system moving at 
velocity $c$: $z=x+ct$, $x-x'=z-z'-c\tau$. The equation for dispersing seeds $u(x,t)$ becomes: 
\begin{multline}
c\,u'(z) = r\, u(z)(k_u-u(z))  + D u''(z) \\
- \alpha \int^{+\infty}_{-\infty} \frac{e^{-\frac{(z-z'-c\tau)^2}{4D\tau}}}{\sqrt{4\pi 
D\tau}}u(z')(k_u-u(z'))dz' 
\label{ecusimp}
\end{multline}
where $r_u$ and the $f$ dependence have been absorbed in the parameters $r$ and $\alpha$ without 
loss of generality, while $u'$ and $u''$ represent first and second derivatives with respect to the 
single variable $z$. This equation cannot be solved analytically, but we can perform a perturbative 
analysis in order to obtain an approximate solution. We can deal with the integral of 
Eq.~(\ref{ecusimp}) making use of Laplace's  formula~\citep[see][]{estrada1994}, which is an asymptotic 
approximation of integrals of the form:
\begin{equation}
I(\lambda)=\int^{b}_{a} e^{-\lambda h(x)}\phi(x)\,dx,
\end{equation}
when $\lambda=(4D\tau)^{-1} \rightarrow \infty$ ($\tau \rightarrow 0$) and $h(x)$ is real. 
Following \citet{estrada1994} and taking $k_u=1$ without loss of generality,
we 
can approximate our integral as:
\begin{multline}
\frac{1}{\sqrt{4D\pi\tau}}I(\lambda) \approx u(z-c\tau)\big(1-u(z-c\tau)\big) \\
	+ D\tau\Big[\big(1-2u(z-c\tau)\big)u''(z-c\tau) \\
	            -2u'(z-c\tau)^2\Big]+\cdots
\end{multline}

\begin{figure}[t]
\centering
\includegraphics[width=\columnwidth, clip=true]{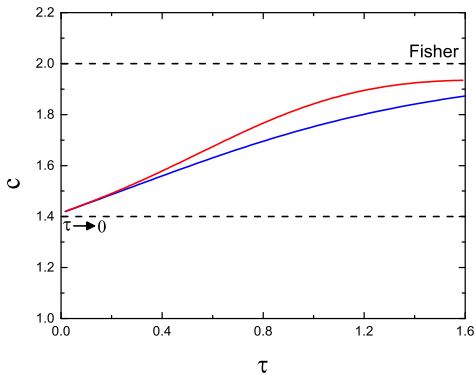}
\caption{Relation between the wave velocity $c$ and the delay $\tau$, obtained by the asymptotic 
(red) and iteration (blue) methods explained in the text. In this case $\alpha=0.5$, $D=1$, $r=1$. 
We also show the special cases of $\tau\to 0$ and $\tau\to\infty$ (Fisher's velocity of the animals' wave), which can be calculated 
exactly.}
\label{figure:c_vs_tau}
\end{figure}

Using the leading terms of the integral expansion in $u$, together with the travelling wave ansatz, 
we finally reduce the problem to the following ordinary differential equation:
\begin{multline}
cu'(z) = Du''(z)+ru(z)\big(1-u(z)\big) \\
	-\alpha\Big(u(z-c\tau)\big(1-u(z-c\tau)\big) \\
	\shoveright{+D\tau\Big[\big(1-2u(z-c\tau)\big)u''(z-c\tau)} \\
	                       -2u'(z-c\tau)^2\Big]\Big).
\end{multline}
We propose an exponential solution $u=A e^{-\lambda z}$ for this equation, which gives rise to a 
transcendental characteristic equation. Considering up to the third order term in $u$ and up to the 
second 
order in a small-$\tau$ expansion, we arrive at the following characteristic equation valid for 
small delays:
\begin{multline}
\alpha D c \tau^2 \lambda^3 + (D-\alpha D \tau - \alpha \frac{c^2}{2} \tau^2) \lambda^2 \\
+\lambda (\alpha c \tau  - c) + r-\alpha=0. \label{caracteristica}
\end{multline}
The relation between the wave velocity $c$ and the delay $\tau$ is provided by the discriminant of 
the solution of Eq.~(\ref{caracteristica}), which we show graphically in Fig.~\ref{figure:c_vs_tau} 
(red curve) (for $\alpha=0.5$, $D=1$ and $r=1$). We can see that there is a growing dependence, 
giving faster waves for larger values of the delay, indicating the velocity enhancement provided by the mobile dispersers. Observe also in Fig.~\ref{figure:c_vs_tau} the 
two limits that can be calculated exactly. The first one corresponds to $\tau\to 0$, when the 
diffusion propagator tends to a Dirac delta $\delta(\tau)$, in which case the integral can be 
evaluated exactly. The second is $\tau\to\infty$, leaving just Fisher's equation in 
Eq.~(\ref{ecu}), 
and then $c=2\sqrt{rD}$. Note that this waves we found
are slower than those corresponding to the movement of the dispersing agent itself, but faster that 
 the limit of immediate deposition of the seeds. We analyse below 
another approximate analytic solution, which provides a similar result.

%----------------------------------------------------------------------------------
\subsection{Iteration method}

Another analytical approach to solve Eq.~(\ref{ecu}) consists in using the iteration technique 
presented in \citep{wu_zou_2001} and \citep{zou_2001}, where the existence of wave-front solutions 
of similar equations is proven, provided that $c$ and $\tau$ satisfy certain relation (see Theorem 
3.2 of \citet{zou_2001}). 

In order to find that relation, we propose again the change of variables $u(x,t)=u(z)=u(x+ct)$. Using 
this method in Eq.~(\ref{ecu}) (with the same specific form of $I$ as in 
Eq.~(\ref{ingestion}), in which $f$ is a parameter), we arrive at the following integro-differential 
equation:
\begin{multline}
c\, u'(z)= r\, u(z) + D\, u''(z) \\- 
\alpha\int^{+\infty}_{-\infty} \frac{e^{-\frac{(z-z'-c\tau)^2}{4D\tau}}}{\sqrt{4\pi D\tau}} 
ru(z')\,dz',
\end{multline}
where we have linearised the reaction terms. 

With the ansatz $u(z)=Ae^{\lambda z}$ we obtain the following characteristic equation:
\begin{equation}
 \Delta(\lambda)\equiv D \lambda^2 - c\lambda + r - \alpha r e^{D \lambda^2 \tau - \lambda c 
\tau}=0.
\end{equation}
The condition to have a single root of this transcendental equation leads to the requirement that: 
\begin{equation}
\tau=\frac{4D}{c^2}\ln\left(\frac{4D\alpha r}{4Dr-c^2}\right). 
\label{tauc}
\end{equation}
Equation (\ref{tauc}) gives a relation between the wave velocity $c$ and the delay $\tau$ that we 
can explore. Figure \ref{figure:c_vs_tau} shows this result (blue curve) for the same values of the 
other parameters as we used for the asymptotic result (red curve). We can see that the two curves 
coincide for small $\tau$, where the asymptotic expansion described above is valid. For larger 
values of $\tau$ the curves separate from each other, as the red curve from the small-$\tau$ expansion loses its 
validity. The iterative result also tends to the right limit as $\tau\to\infty$, albeit 
more slowly. In the following we analyse a numerical solution of the same reduced system. 

%----------------------------------------------------------------------------------
\subsection{Numerical solution}

\begin{figure}[t]
\centering
\includegraphics[width=\columnwidth, clip=true]{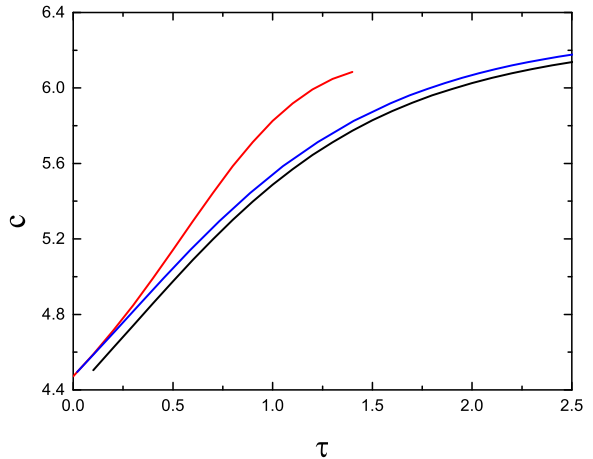}
\caption{Relation between the wave velocity $c$ and the delay $\tau$, obtained by the asymptotic 
(red), iteration (blue) and numerical (black) methods explained in the text. In this case 
$\alpha=0.5$, $D=10$, $r=1$.}
\label{figure:c_vs_tau_numeric}
\end{figure}

The restricted dynamics of $u$---Eq.~(\ref{ecu})---can also be solved numerically. Specifically, we 
solved
\begin{multline}
\frac{\partial u(x,t)}{\partial t} =  r_u u(x,t)\big(1-u(x,t)\big) + D \nabla^2 u(x,t) \\
    - \alpha\! \int^{+\infty}_{-\infty}\! G(x,t|x',t\!-\!\tau)\,r_u u(x',t-\tau) \\
	\big(1-u(x',t-\tau)\big)\,dx',
\end{multline}
where we have absorbed all the dependence on $f$ (which is here a parameter) in the growth rate 
$r_u$, and set $k_u=1$. 
We analysed the formation and propagation of a travelling wave front from an initial stationary 
step 
function:
\begin{equation}
u(x,0)=\begin{cases}1 & \text{if }x<0,\\
                    0 & \text{if }x>0.
		\end{cases}
\end{equation}		
We also set free conditions at the left and right borders of the finite space, and we measure the 
velocity of the solution after a transient time when the front accelerates, and before it reaches 
the borders.

Figure \ref{figure:c_vs_tau_numeric} shows a typical numerical result of the relation between the 
wave velocity and the delay. The plot also shows the corresponding curves obtained by the 
analytical methods. We can see that both of them give good approximations of the wave velocity. In 
particular, the iteration result provides a better approximation for the whole range of delays, 
interpolating well between the limit cases. Both methods provide slightly overestimated velocities.

%----------------------------------------------------------------------------------
\section{Coupled dynamics}

Let us consider the fact that the characteristic time of seed dispersion on the one hand, and the 
time of establishing new fructifying plants on the other, are typically very different. This time 
scale diversity allows a simplified analysis of the coupled model of Eqs.~(\ref{ecf}-\ref{ecs}), 
more complete than the single-component simplification made in the previous section. Essentially, 
this can be accomplished by eliminating the population $s$, i.e. setting the term $r_f\,s(x,t)$ 
(which is the reproduction factor in the definition of the function $F$ in Eq.~(\ref{efe})) equal 
to $r_f\,u(x,t)$. Effectively, this represents an instantaneous germination of the deposited seeds 
into fructifying plants, and the system reduces to:
\begin{align}
\frac{\partial f(x,t)}{\partial t} &= r_f u(k_f-f) - r_u u(k_u-u)\frac{f}{b+f}, \label{2compf}\\
\frac{\partial u(x,t)}{\partial t} &=  r_u u(k_u-u)\frac{f}{b+f} + D \nabla^2 u(x,t) \notag\\
                                   &~~- \alpha\! \int^{+\infty}_{-\infty}\! 
G(x,t|x',t\!-\!\tau)\,r_u u(x',t-\tau) 
\notag\\
&~~~~\big(k_u-u(x',t-\tau)\big)\frac{f(x',t-\tau)}{b+f(x',t-\tau)}\,dx'. \label{2compu}
\end{align}

Even if the use of an instantaneous germination seems too crude an approximation, it helps to keep the equations tractable, as a different assumption would 
introduce a new delay term. Moreover, we have found that numerical results with an additional germination rate are qualitatively the same as the ones shown here.

Based on the existence and the properties of travelling waves of isolated dispersing seeds found in 
the previous sections, we have searched for the equivalent dynamics in this coupled model. We have 
analysed the system only numerically in this case, with two different initial conditions, which we 
call homogeneous and heterogeneous depending on the initial values of $f(x)$:
\begin{multline}
\text{Homogeneous initial condition: } \\
\begin{cases}
  f(x,0) = k_f, \\
  u(x,0) = k_u~~\text{if }x\le0,~~0~~\text{if }x>0.
\end{cases}
\end{multline}
\begin{multline}
\text{Heterogeneous initial condition: }\\
\begin{cases}
  f(x,0) = k_f~~\text{if }x\le0,~~0~~\text{if }x>0, \\
  u(x,0) = k_u~~\text{if }x\le0,~~0~~\text{if }x>0.
\end{cases}
\end{multline}		  

\begin{figure}[t]
\centering
\includegraphics[width=\columnwidth, clip=true]{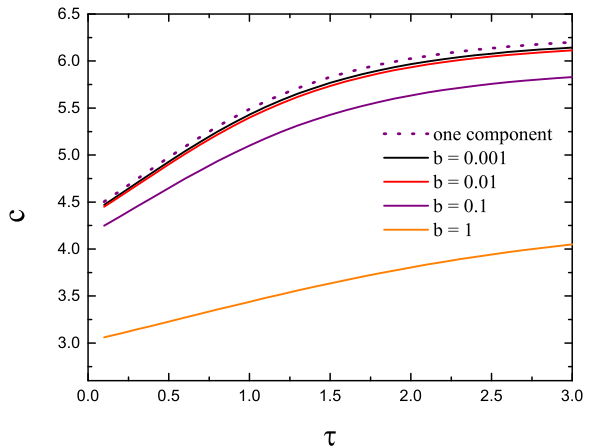}
\caption{Relation between the wave velocity $c$ and the delay $\tau$, obtained numerically for the 
two-components model. $\alpha=0.5$, $D=10$, $r=1$, $b$ as shown. The corresponding result for the one-component model is shown for comparison.}
\label{figure:homo}
\end{figure}

The homogeneous initial condition evolves, after a transient, to a travelling front of the $u$ 
variable, similar to the single component simplified model analysed before. It is accompanied by a 
shallow depletion of $f$ that moves at the same speed. In this regard, it is similar to a case 
where 
$f$ is just a parameter, and we used it as a benchmark to compare the role of the satiation 
parameter $b$ that appears in Eqs.~(\ref{2compf}-\ref{2compu}). The resulting velocity as a 
function 
of $\tau$ is shown in Fig.~\ref{figure:homo}, for different values of $b$. When $b\to 0$ the 
satiation term disappears and the dependence of ingestion on $f$ becomes effectively a parameter, 
just like in the one-component model. For larger values of $b$ the curves show the same shape, with 
the velocity growing with $\tau$ but with smaller velocities for each value of the delay.

\begin{figure}[t]
\centering
\includegraphics[width=\columnwidth, clip=true]{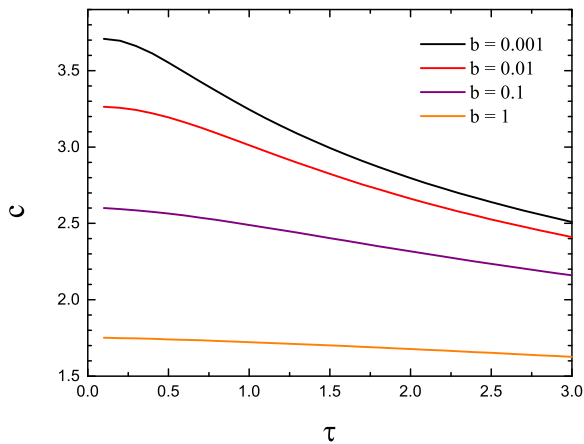}
\caption{Relation between the wave velocity $c$ and the delay $\tau$, obtained numerically for the 
two-components model and heterogeneous initial conditions. $\alpha=0.5$, $D=10$, $r=1$, $b$ as 
shown.}
\label{figure:hetero}
\end{figure}

The heterogeneous initial condition is the most interesting case in the analysis of the expansion 
of 
a bound patch of vegetation, facilitated by the dispersing agents. For example, one can expect a 
double invasion wave of dispersing seeds and plants, corresponding to the advance of the patch edge. In this case, since $u$ is propagating into empty space, the velocity of the invasion should 
be slower than in the single component and in the two-components with homogeneous initial 
conditions. Our results show that indeed such is the case. The two waves propagate asymptotically 
with the same velocity and a small lag of $f$ behind $u$. In addition we find that the 
dependence of the velocity on the delay is reversed with respect to the one corresponding to those 
cases. Figure~\ref{figure:hetero} shows the results corresponding to the same parameters as those 
used in Fig.~\ref{figure:homo}. 

The reversal of the dependence of the velocity on the delay for different initial conditions can be 
understood in the following way. When the initial vegetation extends homogeneously a longer 
deposition time allows the front of dispersing animals to reach farther and expand faster, since 
their resource $f$ is available wherever they go. On the contrary, when the initial vegetation is 
bounded, the animals cannot reach as far because of the limited resource. In this case, if they 
diffuse farther, the seeds are lost. The border of the region occupied by $u$ needs to propagate 
slower in order to provide for the establishment of their resource. 

Finally, we have analysed the lag between the two fronts and its dependence on the parameters. 
We found that it is almost insensible to the value of $\tau$, but depends strongly on the effective 
reproduction rate $r_f$. Larger values of this parameter can reduce the separation between the 
fronts to almost zero. 

%\begin{figure}[htp]
%\centering\includegraphics[width=\columnwidth, clip=true]{lag_vs_rf.eps}
%\caption{}
%\label{figure:lag}
%\end{figure}

%----------------------------------------------------------------------------------
\section{Discussion and conclusions}

The inconsistency between  migration and estimated invasion rates of plants that would account 
for the dynamics of their population during the Holocene post-glacial migration has been named 
Reid's paradox. Among other plausible explanations, several authors have suggested the 
occurrence of long-distance transport events mediated by animal dispersal. 
Based on this assumptions, we have analysed a model of plant propagation by diffusive dispersion of 
their seeds, mediated by animal ingestion and transport. The model involves three populations: 
immobile seeds in fruits, mobile seeds in animals, and seeds (again immobile) deposited in the 
substrate. In particular, we studied the propagation of travelling waves in the form of invasion 
fronts, arising from Heaviside-like initial conditions. The mathematical problem is similar to that 
of reaction-diffusion waves developed since the 1930s, like the ones studied by Fisher in the 
context of the propagation of a genetic trait in a population \citep{volpert2009}. Like those, the 
velocity of propagation depends on the diffusion constant and on the linear growth parameter of the 
field being dispersed. 

Several authors have previously contributed to the seek of a mathematical formulation of  
zoochory dynamics. \citet{neupane2015b} and \citet{powell04}, for example, model the deposition of dispersed seeds 
with a spatial kernel in an integro-differential (or difference, as in \cite{neubert1995}) 
dynamics. These kernels are then fit from data to compare predictions with field observations. The 
existence of these spatial kernels naturally arises from the time that the seeds travel along with their dispersers (such as 
the gut transit time, as modelled numerically by \cite{morales2006}). The fact that there is a delay 
between the ingestion of the seeds and their deposition at a 
different place provides, from the point of view of the seed, faster fronts than those 
corresponding 
to a negligible delay. This corresponds to a possible resolution of Reid's paradox: a faster 
propagation thanks to the mediation of animal dispersers. In our model the diffusion coefficient of 
the animals, together with the gut transit time of the seeds (or its equivalent delay in other 
transport mechanisms), determine the velocity of propagation of the vegetation front. Our approach is not focused on the 
phenomena that may arise due to spatial structures but on the effect induced by the inherent delay 
associated to an active transportation by an animal. In that sense, all these approaches are complementary.

We have analysed the model through different approximations, which allowed the characterization of 
the phenomenon in one spatial dimension. First we studied a one-component simplification, in which 
only the diffusing seeds remain as dynamical variables. This scenario, in which the density of plants remains constant, may correspond to systems in which the plants are long-lived (e.g. trees). The 
corresponding delayed-integro-differential equation was solved 
for travelling waves using an asymptotic expansion and an iteration procedure. Both methods 
provided the velocity of the waves as a function of the delay parameter. Numerical solution of the 
equation confirmed the validity of the analytic procedures. 
Additionally, we observed that the shape of the front is different from the corresponding to a case 
without delay. Actually, the leading front of the wave has the same properties as the Fisher's 
one---allowing for the calculation of the asymptotic velocity. The main difference is in the 
trailing part of the front, where the invasion tends to the carrying capacity, whose shape depends 
on the delay.

In connection with these findings, we also analysed an intermediate simplification of the model 
consisting of two dynamical 
variables: the fructifying plants and the animals. This situation may be relevant for systems with
annual plants and dispersers that have longer generation times. We found that the general phenomenon (faster 
fronts of plants with respect to no delays, mediated by the animals) is maintained, but there is a 
strong dependence on 
the initial conditions. The propagating wave becomes limited not by diffusion, but by the limited 
resources at the leading edge. As a result of this, the dependence of the velocity of the front on 
the relevant parameter measuring the delayed transport of the seeds, $\tau$, becomes inverted 
(decaying) with respect to the single component model. In this context, it is worth mentioning that 
the behaviour of animals across a patch edge, based on habitat preference, may be relevant in the 
dynamics of the coupled populations, as studied via reaction-diffusion models by 
\citet{maciel2013,maciel2014}. This is a matter to be considered in forthcoming work.

The present analysis does not exhaust the characterization of the solutions of the system. In 
particular, we have studied a single phenomenon: the speed of an invasion front arising from an 
heterogeneous initial condition. While this is certainly relevant for two-species invasions and 
restoration of ecosystems, there are other dynamically interesting problems with distributed 
heterogeneities, such as the irregularity of the topology, of the diffusion coefficient 
\citep{neupane2015b} or of the distribution of resources. We also expect 
to extend it to the study of the three-component waves elsewhere. Besides, in the present work we 
characterized the movement of the disperser as a diffusive 
phenomenon. Other transport mechanisms could be considered once a persistent interaction between 
the 
plant and the animal has been established and the topology of the landscape has been shaped. In 
those cases, we could add chemotactic terms or Cahn-Hilliard like equations \citep{liu2016} to take 
into 
consideration the feedback interactions between both species. 

\section*{Acknowledgements}
This work received support from the Consejo Nacional de Investigaciones Cient\'{\i}ficas y 
T\'ecnicas (PIP 112-201101-00310), 
Universidad Nacional de Cuyo (06/C304). We also thank Javier Fern\'{a}ndez and Juan Manuel Morales for 
fruitful discussions.

\end{document}